Graphical Abstract

# A Diffusion Approximation model of Active Queue Management

Dariusz Marek, Adam Domański, Joanna Domańska, Tadeusz Czachórski, Jerzy Klamka, Jakub Szyguła

# Highlights

## A Diffusion Approximation model of Active Queue Management

Dariusz Marek, Adam Domański, Joanna Domańska, Tadeusz Czachórski, Jerzy Klamka, Jakub Szyguła

- Dariusz Marek, Ph.D. Student in the Institute of Computer Science, Faculty of Automatic Control, Electronics and Computer Science, Silesian University of Technology.

- Adam Domański, Ph.D., Eng., works in the Computer Equipment Group of the Institute of Computer Science, Faculty of Automatic Control, Electronics and Computer Science, Silesian University of Technology. His main research interest in the computer networks domain is congestion control in packet networks.

- Joanna Domańska, Ph.D., Eng., works in the Computer Systems Modelling and Performance Evaluation Group of the Institute of Theoretical and Applied Informatics, Polish Academy of Sciences. Her main areas of research include performance modeling methods for computer networks.

- Tadeusz Czachórski, Ph.D., Prof., the head of the Institute of Theoretical and Applied Informatics of the Polish Academy of Sciences. His main areas of interest are mathematical and numerical methods for modeling and evaluation of computer networks.

- Jerzy Klamka, Ph.D., Prof., a full member of the Polish Academy of Sciences, works in the Quantum Systems of Informatics Group of the Institute of Theoretical and Applied Informatics of the Polish Academy of Sciences. His main areas of research are controllability and observability of linear and nonlinear dynamical systems, and mathematical foundations of quantum computations. He is an author of monographs and numerous papers published in international journals.

- Jakub Szyguła, Ph.D. Student in the Institute of Computer Science, Faculty of Automatic Control, Electronics and Computer Science, Silesian University of Technology.

# A Diffusion Approximation model of Active Queue Management[⋆],[⋆⋆]

Dariusz Marek[a,1], Adam Domański[a,2], Joanna Domańska[b,3], Tadeusz Czachórski[b,4], Jerzy Klamka[b,5] and Jakub Szyguła[a,6]

[a]*Institute of Informatics, Silesian University of Technology, Address: Akademicka 16, 44-100 Gliwice, Poland*
[b]*Institute of Theoretical and Applied Informatics, Polish Academy of Sciences, Address: ul. Bałtycka 5, 44-100 Gliwice, Poland*

ARTICLE INFO

ABSTRACT

The article describes the diffusion approximation and the method of its use for evaluation of the effectiveness of active queue management (AQM) mechanisms. The presented model combines the approximation and simulation approaches. The diffusion is used to estimate queue distributions. The goal of the simulation part of the model is the evaluation of the AQM mechanism. Based on the obtained queue distributions, the simulation part of the model decides on package losses and influences the flow intensity of the transmitter. The performance of fractional order $PI^\alpha$ controller is compared to the performance of RED, a well known active queue management mechanism.

## 1. Introduction

The efficiency of the TCP protocol depends largely on the queue management algorithm used in a router. There are distinguished two basic principles of network packet management. In passive queue management, packets coming to a buffer are rejected only if the buffer is fully occupied. The active mechanisms are based on the preventive packet dropping when there is still place to store them in the queue. The packets are dropped randomly and usually in accordance with the assumed probability loss function. This approach enhances the throughput and fairness of the link sharing and eliminates global synchronization. The basic active queue management algorithm is Random Early Detection (RED), primarily proposed in 1993 by Sally Floyd and Van Jacobson [15]. There were several works studying the impact of various parameters on the RED performance [2], [26] and many variations of AQM mechanism were developed to improve its performance, e.g. [41], [7].

### 1.1. Our contribution

This paper enhances the $PI^\alpha$ controller performance with the use of diffusion approximation analysis and simulation. We propose a model of a TCP connection having a bottleneck router with AQM policy. The proposed model is a combination of diffusion and simulation model. We use diffusion approximation to estimate the distribution of the queue length. On the basis of the distribution, the average queue length is calculated.

Depending on the computed average queue length, the simulation part of a model makes the decision about the packet rejection. In the case of active queue management, this decision of dropping packets is random and depends on queue occupancy and dropping packet probability function. The decision of packet rejection halves the intensity $\lambda$ of the traffic source. In the absence of a loss, the intensity of the source increases linearly. This mechanism may be seen as a part of the closed loop control of TCP/IP traffic intensity. The evolution of the source intensity can be illustrated as follows:

$$\lambda = \begin{cases} \lambda + \zeta & \text{if} \quad AQM\ decides\ no\ loss \\ \frac{\lambda}{2} & \text{if} \quad AQM\ decides\ a\ loss \end{cases}$$

where $\zeta$ is assumed a constant increase in the intensity of the source. Such source behavior is a simplified model of the TCP NewReno algorithm [14]. Together with the use of RED algorithm, the decision of a packet rejection is also given here by a $PI^\alpha$ controller.

This article investigates the behavior of fractional order controller $PI^\alpha$ applied to control Internet traffic supervised by TCP transport protocol. We investigate the influence of parameters of the controller on the queue length and evolution of the TCP congestion window. The performance of fractional order $PI^\alpha$ controller is compared to the performance of RED, a well known active queue management mechanism.

The remainder of the paper is organized as follows: Section 2 reminds works related to this article. Section 3 summarizes the basics of the diffusion approximation. In this section we present two used networks station models: unlimited queue: G/G/1 and limited queue: G/G/1/N. Section 4 presents theoretical bases for $PI^\alpha$ controller, next used in proposed mixed diffusion approximation-simulation model. Section 5 discusses the obtained results. Concluding remarks are presented in section 6

## 2. Related works

Considerations for use of the diffusion approximation and the G/G/1 queue for the traffic server performance are presented in [34]. Some works developed transient state analysis based on diffusion and fluid flow approximation [6] and

[⋆]This research was partially financed by National Science Center project no. 2017/27/B/ST6/00145.
[⋆]Corresponding author
[⋆⋆]Principal corresponding author
ORCID(s):
[1]0000-0002-9452-8361
[2]0000-0002-1935-8358
[3]0000-0001-7158-0258
[4]0000-0003-1574-9826
[5]0000-0002-4074-8502
[6]0000-0002-5728-6105

studied also priority queues [4], [5]. Proposed in this paper solution is also based on diffusion approximation and G/G/1/N queuing model like in [38]. The increase of the number of users with permanent access to the Internet has created renewed interest in the research of multi-server queuing models. In the article [17] the limits of the GI/M/n/∞ were studied.

The initial AQM mechanism RED (Random Early Detection) was proposed by IETF and was primarily described by Sally Floyd and Van Jacobson in [15]. It is based on a drop function giving the probability that a packet is rejected. The argument of this function is a weighted moving average queue length, acting as a low-pass filter. This average depends on actual queue occupancy and a previously calculated value of the weighted moving average. The packet dropping probability is based on a linear function.

The introduction of AQM mechanisms has significantly improved the quality of network transmission. However, RED has such drawbacks as low throughput, unfair bandwidth sharing, introduction of variable latency, deterioration of network stability [18]. For these reasons, numerous propositions of improvements appeared. One of these modifications is DSRED (double-slope RED) introduced and developed in [41] [42]. This solution resolves the linear packet dropping function by the drop function composed of two lines with different slopes. The paper [43] proposes the NLRED algorithm with a quadratic dropping function. In the paper [1] authors considered the polynomial function.

The RED mechanism may be also replaced by any other controller. The article [19] describes a Proportional-Integral controller on the low-frequency dynamics. The paper [36] described a robust controller, based on a known technique for $H^\infty$ control of systems with time delays. The $PI^\alpha$ AQM controller proposed in [19] was designed following the small-gain theorem. Easy implementation of $PI^\alpha$ AQM controllers in real networks resulted in a number of propositions [29], [27], [39], [28]. The first application of the fractional order $PI^\alpha$ controller as an AQM policy in a fluid flow model of a TCP connection was presented in [25].

## 3. Diffusion approximation

The main principle of the diffusion approximation method [30], [31], [32], [33] is replacing the discrete process $N(t)$ by a continuous diffusion process $X(t)$. The changes of $dX(t) = X(t+dt) - X(t)$ are normally distributed with the mean $\beta dt$ and variance $\alpha dt$.

The diffusion approximation method presents the equation:

$$\frac{\partial f(x,t;x_0)}{\partial t} = \frac{\alpha}{2}\frac{\partial^2 f(x,t;x_0)}{\partial x^2} - \beta \frac{\partial f(x,t;x_0)}{\partial x} \quad (1)$$

where $\beta$, $\alpha$ are parameters of the diffusion equation and $f(x)$ is the density of the diffusion process. The solution of Eq. (1), the approximation can be used for the behavior of a queue evaluation having interarrival times following any distribution $A(x)$ stream with any service time distribution $B(x)$. In the case of diffusion approximation two first moments of this distributions are considered: $E[A] = 1/\lambda$, $E[B] = 1/\mu$, $Var[A] = \sigma_A^2$, $Var[B] = \sigma_B^2$. Denote also of variation: $C_A^2 = \sigma_A^2 \lambda^2$, $C_B^2 = \sigma_B^2 \mu^2$.

The probability density function (pdf) of this process present formula [16]:

$$f(x,t;x_0)\,dx = P[x \leq X(t) < x+dx|X(0) = x_0] \quad (2)$$

approximates the distribution of the number of customers in the investigated system. To ensure an increase in distribution with the same frequency as the length of the period of the observation, the diffusion coefficients should be set as follows:

$$\beta = \lambda - \mu, \quad \alpha = \sigma_A^2 \lambda^3 + \sigma_B^2 \mu^3 = C_A^2 \lambda + C_B^2 \mu \quad (3)$$

### 3.1. Unlimited queue: G/G/1 station

More formal justification of diffusion approximation is in limit theorems for $G/G/1$ system given by Iglehart and Whitte [21, 22, 40, 20]. If $\hat{N}_n$ is a series of random variables derived from $N(t)$:

$$\hat{N}_n = \frac{N(nt) - (\lambda - \mu)nt}{(\sigma_A^2 \lambda^3 + \sigma_B^2 \mu^3)\sqrt{n}} \quad (4)$$

then the series is weakly convergent (in the sense of distribution) to $\xi$ where $\xi(t)$ is a standard Wiener process (i.e. diffusion process with $\beta = 0$ i $\alpha = 1$) provided that $\varrho > 1$, that means if the system is overloaded and has no equilibrium state. In the case of $\varrho = 1$ the series $\hat{N}_n$ is convergent to $\xi_R$. The $\xi_R(t)$ process is $\xi(t)$ process limited to half-axis $x > 0$:

$$\xi_R(t) = \xi(t) - \inf[\xi(u), \ 0 \leq u \leq t]. \quad (5)$$

The process $N(t)$ is never negative. $X(t)$ should also be limited to value $x \geq 0$. The method of solving this limitation is placing *reflecting barrier* [24] at point $x = 0$. The barrier limits the process to a positive $x$-axis:

$$\int_0^\infty f(x,t;x_0)dx = 1 \quad (6)$$

and:

$$\frac{\partial}{\partial t}\int_0^\infty f(x,t;x_0)dx = \int_0^\infty \frac{\partial f(x,t;x_0)}{\partial t}dx = 0. \quad (7)$$

Replacing the integrated function with the right side of the diffusion equation we get the boundary condition corresponding to the reflecting barrier at zero:

$$\lim_{x \to 0}\left[\frac{\alpha}{2}\frac{\partial f(x,t;x_0)}{\partial x} - \beta f(x,t;x_0)\right] = 0. \quad (8)$$

The solution of Eq. (2) with boundary conditions defined by Eq. (8) gives us: [24]

$$f(x,t;x_0) = \frac{\partial}{\partial x}\left[\Phi\left(\frac{x-x_0-\beta t}{\alpha t}\right) - e^{\frac{2\beta x}{\alpha}}\Phi\left(\frac{x+x_0+\beta t}{\alpha t}\right)\right], \quad (9)$$

where: $\Phi(x) = \int_{-\infty}^{x} \frac{1}{\sqrt{2\pi}} e^{-t^2/2} dt$ is the PDF of standard normal distribution.

For $(\beta < 0)$ the system converges to a steady-state.

$$\lim_{t \to \infty} f(x, t; x_0) = f(x) \quad (10)$$

and partial differential equation (1) becomes an ordinary one:

$$0 = \frac{\alpha}{2} \frac{d^2 f(x)}{dx^2} - \beta \frac{d f(x)}{dx} \quad (11)$$

with solution:

$$f(x) = -\frac{2\beta}{\alpha} e^{\frac{2\beta x}{\alpha}}.$$

This formula approximates the queue at $G/G/1$ system:

$$p(n, t; n_0) \approx f(n, t; n_0), \quad (12)$$

and at steady-state [23] $p(n) \approx f(n)$; one can also choose e.g.

$$p(0) \approx \int_0^{0.5} f(x)dx, p(n) \approx \int_{n-0.5}^{n+0.5} f(x), n = 1, 2, \ldots, \quad (13)$$

The reflecting barrier in point $x = 0$ causes the immediate reflecting of the process ("zero" value is excluded from the process). The value of $f(0) = 0$. This version of the diffusion process is a heavy-load approximation and gives reasonable results in case of a system with utilization factor $\rho$ close to 1.

This difficulty removes another limit condition at $x = 0$: a *barrier with instantaneous elementary jumps* [16]. When the diffusion process reaches $x = 0$, it returns to $x = 1$ (it corresponds to the arrival of one packet to the system). The time when the process is at $x = 0$ means that the system is in the idle state. The diffusion equation becomes [16]:

$$\frac{\partial f(x, t; x_0)}{\partial t} = \frac{\alpha}{2} \frac{\partial^2 f(x, t; x_0)}{\partial x^2}$$
$$-\beta \frac{\partial f(x, t; x_0)}{\partial x} + \lambda p_0(t) \delta(x - 1), \quad (14)$$

$$\frac{dp_0(t)}{dt} = \lim_{x \to 0} \left[\frac{\alpha}{2} \frac{\partial f(x, t; x_0)}{\partial x} - \beta f(x, t; x_0)\right] - \lambda p_0(t),$$

where: $p_0(t) = P[X(t) = 0]$.

The term $\lambda p_0(t) \delta(x-1)$ gives the probability density that the process is started at point $x = 1$ at the moment $t$, because of the jump from the barrier. The second equation balances $p_0(t)$: the term $\lim_{x \to 0}[\frac{\alpha}{2} \frac{\partial f(x, t, x_0)}{\partial x} - \beta f(x, t; x_0)]$ gives the probability flow *into* the barrier, while $\lambda p_0(t)$ represents the probability flow *out* of the barrier.

The approach to obtain the function $f(x, t; x_0)$ of the process with jumps from the barrier, see [38], is to express it with the use of another pdf $\phi(x, t; x_0)$ for the diffusion process with the absorbing barrier at $x = 0$ [38]. This process starts at $t = 0$ from $x = x_0$ and ends when it reaches the barrier. Such probability density function is easier to determine [3],

$$\phi(x, t; x_0) = \frac{e^{\frac{\beta}{\alpha}(x-x_0) - \frac{\beta^2}{2\alpha}t}}{\sqrt{2\pi\alpha t}} [e^{-\frac{(x-x_0)^2}{2\alpha t}} - e^{-\frac{(x+x_0)^2}{2\alpha t}}] \quad (15)$$

The density function of the first step time from $x = x_0$ to $x = 0$ is:

$$\gamma_{x_0,0}(t) = \lim_{x \to 0} \left[\frac{\alpha}{2} \frac{\partial}{\partial x} \phi(x, t; x_0)\right.$$

$$\left. -\beta \phi(x, t; x_0)\right] = \frac{x_0}{\sqrt{2\pi\alpha t^3}} e^{-\frac{(\beta t+1)^2}{2\alpha t}} \quad (16)$$

The process starts at $t = 0$ at a point $x$ with density $\psi(x)$ and every time, when it reaches the barrier, it stays there for a time defined by a density function $l_0(x)$ and then jump to the point $x = 1$. The total stream $\gamma_0(t)$ of probability mass that enters the barrier is:

$$\gamma_0(t) = p_0(0)\delta(t) + [1-p_0(0)]\gamma_{\psi,0}(t) + \int_0^t g_1(\tau)\gamma_{1,0}(t-\tau)d\tau, \quad (17)$$

where:

$$\gamma_{\psi,0}(t) = \int_0^\infty \gamma_{\xi,0}(t)\psi(\xi)d\xi, \quad (18)$$

$$g_1(\tau) = \int_0^\tau \gamma_0(t)l_0(\tau - t)dt. \quad (19)$$

The density function of the diffusion process with instantaneous returns is:

$$f(x, t; x_0) = \phi(x, t; \psi) + \int_0^t g_1(\tau)\phi(x, t-\tau; 1)d\tau. \quad (20)$$

For the Eqs. (17) and (19) there are following Laplace transform equations:

$$\bar{\gamma}_0(s) = p_0(0) + [1 - p_0(0)]\bar{\gamma}_{\psi,0}(s) + \bar{g}_1(s)\bar{\gamma}_{1,0}(s),$$

$$\bar{g}_1(s) = \bar{\gamma}_0(s)\bar{l}_0(s) \quad (21)$$

where:

$$\bar{\gamma}_{x_0,0}(s) = e^{-x_0 \frac{\beta + A(s)}{\alpha}} \quad (22)$$

$$\overline{\gamma}_{\psi,0}(s) = \int_0^\infty \overline{\gamma}_{\xi,0}(s)\psi(\xi)d\xi \qquad (23)$$

and then:

$$\overline{g}_1(s) = [p_0(0) + [1-p_0(0)]\overline{\gamma}_{\psi,0}(s)] \frac{\overline{l}_0(s)}{1-\overline{l}_0(s)\overline{\gamma}_{1,0}(s)} \qquad (24)$$

Equation (20) in terms of Laplace transform becomes:

$$\overline{f}(x, s; x_0) = \overline{\phi}(x, s; \psi) + \overline{g}_1(s)\overline{\phi}(x, s; 1), \qquad (25)$$

where:

$$\overline{\phi}(x, s; x_0) = \frac{e^{\beta(x_C x_0)}}{A(s)} [e^{-|x-x_0|\frac{A(s)}{\alpha}} - e^{-|x+x_0|\frac{A(s)}{\alpha}}] \qquad (26)$$

$$\overline{\phi}(x, s; \psi) = \int_0^\infty \overline{\phi}(x, s; \xi)\psi(\xi)d\xi, \ A(s) = \sqrt{\beta^2 + 2\alpha s}. \qquad (27)$$

The inverse transforms of these functions can be obtained numerically. For this purpose, Stehfest's algorithm [37] is used. Our researches have shown that the results obtained by other methods deviate significantly from the correct results.

The presented above G/G/1 model assumes constant model parameters. In the case of variable model parameters, we should divide the model time into time periods, where the parameters of the model are constant. In this approach the results obtained at the end of one time period serves as the initial condition for the next period.

### 3.2. Limited queue: G/G/1/N station

In the case of G/G/1/N station, the number of packets in the node is limited to $N$. The implementation of such an assumption is based on the second barrier placed at $x = N$. So G/G/1/N model is limited with two barriers. The first barrier is placed in $x = 0$ and the second in place at $x = N$. The behavior of the process in the first barrier we described in the previous subsection. In the second barrier when the process reaches $x = N$, it waits in this state some time, this time corresponding to the period when the queue is full and incoming packets are being lost plus the time of realization of forwarding the current packets, and then jumps to $x = N-1$.

The density function of such process $f(x, t; x_0)$ is obtained similarly to described above model with infinity queue.

In the first step of the method, it determines the density $\phi(x, t; x_0)$ of the diffusion process with two absorbing barriers at $x = 0$ and $x = N$, started at $t = 0$ from $x = x_0$ [3]:

$$\phi(x, t; x_0) = \frac{1}{\sqrt{2\pi\alpha t}} \sum_{n=-\infty}^{\infty} (\exp[\frac{\beta x'_n}{\alpha} - \frac{(x-x_0-x'_n-\beta t)^2}{2\alpha t}]$$

$$- \exp[\frac{\beta x''_n}{\alpha} - \frac{(x-x_0-x''_n-\beta t)^2}{2\alpha t}]), \qquad (28)$$

where: $x'_n = 2nN, x''_n = -2x_0 - x'_n$.

For the initial condition is defined by a function $\psi(x)$, $x \in (0, N)$, $\lim_{x\to 0}\psi(x) = \lim_{x\to N}\psi(x) = 0$, then the pdf of the process has the form:

$$\phi(x, t; \psi) = \int_0^N \phi(x, t; \xi)\psi(\xi)d\xi. \qquad (29)$$

Diffusion process function $f(x, t; \psi)$ with instantaneous returns from both barriers is composed of function $\varphi(x, t; \psi)_1$ representing the influence of the initial condition and of the set of functions $\varphi(x, t-\tau; 1)$, $\varphi(x, t-\tau; N-1)$ started after the jump from barrier at time $\tau < t$, at points $x = 1$ and $x = N - 1$ with intensities $g_1(\tau)$ and $g_{N-1}(\tau)$ [38]:

$$f(x, t; \psi) = \varphi(x, t; \psi) + \int_0^t g_1(\tau)\varphi(x, t-\tau; 1)d\tau$$

$$+ \int_0^t g_{N-1}(\tau)\varphi(x, t-\tau; N-1)d\tau. \qquad (30)$$

Functions $g_1(t), g_N(t)$ can be expressed by means of probability densities $\gamma_0(t)$ and $\gamma_N(t)$:

$$g_1(\tau) = \int_0^\tau \gamma_0(t)l_0(\tau-t)dt,$$

$$g_{N-1}(\tau) = \int_0^\tau \gamma_N(t)l_N(\tau-t)dt, \qquad (31)$$

where: $l_0(x)$, $l_N(x)$ are distribution density functions of time, when process stays at $x = 0$ and $x = N$.

Probability densities $\gamma_0(t), \gamma_N(t)$, that process enters the barrier at $x = 0$, or $x = N$ at time $t$ are:

$$\gamma_0(t) = p_0(0)\delta(t) + [1 - p_0(0) - p_N(0)]\gamma_{\psi,0}(t)$$

$$+ \int_0^t g_1(\tau)\gamma_{1,0}(t-\tau)d\tau + \int_0^t g_{N-1}(\tau)\gamma_{N-1,0}(t-\tau)d\tau, \qquad (32)$$

$$\gamma_N(t) = p_N(0)\delta(t) + [1 - p_0(0) - p_N(0)]\gamma_{\psi,N}(t)$$

$$+ \int_0^t g_1(\tau)\gamma_{1,N}(t-\tau)d\tau + \int_0^t g_{N-1}(\tau)\gamma_{N-1,N}(t-\tau)d\tau, \qquad (33)$$

where: $\gamma_{1,0}(t), \gamma_{1,N}(t), \gamma_{N-1,0}(t), \gamma_{N-1,N}(t)$ are the densities of first passage times of diffusion process between corresponding points (indicated in the index), e.g.:

$$\gamma_{1,0}(t) = \lim_{x\to 0}[\frac{\alpha}{2}\frac{\partial\phi(x,t;1)}{\partial x} - \beta\phi(x,t;1)]. \qquad (34)$$

Functions $\gamma_{\psi,0}(t)$, $\gamma_{\psi,N}(t)$ define probability densities, that the initial process, started on $t = 0$, at the point $x = \xi$ with density $\psi(\xi)$, will end at time $t$ by entering the barrier respectively at $x = 0$ or $x = N$.

Densities $g_1(t)$ and $g_N(t)$ may be expressed with the use of functions $\gamma_0(t)$ and $\gamma_N(t)$:

$$g_1(\tau) = \int_0^\tau \gamma_0(t) l_0(\tau - t) dt,$$

$$g_{N-1}(\tau) = \int_0^\tau \gamma_N(t) l_N(\tau - t) dt, \qquad (35)$$

where: $l_0(x)$, $l_N(x)$ are the densities of sojourn times in $x = 0$ and $x = N$, · the distributions of these times are not restricted to exponential ones. Finally, probabilities, that at time $t$ process has value $x = 0$, or $x = N$ are:

$$p_0(t) = \int_0^t [\gamma_0(\tau) - g_1(\tau)] d\tau,$$

$$p_N(t) = \int_0^t [\gamma_N(\tau) - g_{N-1}(\tau)] d\tau. \qquad (36)$$

Similarly to the previous model, the presented equations are transformed using Laplace transform and determine the transformed function $\overline{f}(x, s; \psi)$ and it's original is calculated numerically. To vary with time the diffusion coefficients, we divide the time axis into subintervals with specific constant parameters and the solution of the end of an interval gives initial conditions for the next one. The presented transient solution tends for $t \to \infty$, to the known steady-state solutions thus models based on diffusion approximation can also refer to the steady-state.

## 4. AQM mechanism based on non-integer order $PI^\alpha$ controller.

Fractional Order Derivatives and Integrals (FOD/FOI) are a natural extension of the well-known integrals and derivatives. Differintegrals of non-integer orders enable better and more precise control of physical processes. A proportional-integral controller (PI controller) is a traditional mechanism used in feedback control systems. Earlier works show that the non-integer order controllers have better behavior than classic controllers [35].

The articles [9], [8], [12], [10] describe how to use the response of the $PI^\alpha$(non-integer integral order) to calculate the response of the AQM mechanism. The probability of packet loss is given by the formula:

$$p_i = max\{0, -(K_P e_k + K_I \Delta^\alpha e_k)\} \qquad (37)$$

where $K_P$, $K_I$ are tuning parameters, $e_k$ is the error in current slot $e_k = Q_k - Q$, i.e. the difference between current queue $Q_k$ and desired queue $Q$.

Thus, the dropping probability depends on three parameters: the coefficients for the proportional and integral terms $K_p$, $K_i$ and integral order $\alpha$.

In the active queue management, packet drop probabilities are determined at discrete moments of packet arrivals, hence the queue model should be considered as a case of discrete systems. There is only one definition of the discreet differ-integrals of non-integer order. This definition is a generalization of the traditional definition of the difference of integer order to the non-integer order and it is analogous to a generalization used in Grunwald-Letnikov (GrLET) formula.

For a given sequence $f_0, f_1, ..., f_j, ..., f_k$

$$\triangle^\alpha f_k = \sum_{j=0}^{k} (-1)^j \binom{\alpha}{j} f_{k-j} \qquad (38)$$

where $\alpha \in R$ is generally a non-integer fractional order, $f_k$ is a differentiated discrete function, and $\binom{\alpha}{j}$ is generalized Newton symbol defined as follows:

$$\binom{\alpha}{j} = \begin{cases} 1 & \text{for } j = 0 \\ \dfrac{\alpha(\alpha - 1)(\alpha - 2)..(\alpha - j + 1)}{j!} & \text{for } j = 1, 2, \ldots \end{cases} \qquad (39)$$

For $\alpha = 1$ we get the formula for the difference of the first order (only two coefficients are non-zero).

$$\triangle^1 x_k = 1x_k - 1x_{k-1} + 0x_{k-2} + 0x_{k-3} \ldots \qquad (40)$$

For $\alpha = -1$ we get the sum of all samples (the discrete integral of first order equivalent).

$$\triangle^{-1} x_k = 1x_k + 1x_{k-1} + 1x_{k-2} + 1x_{k-3} \ldots \qquad (41)$$

For non-integer derivative and integral order we get the weighted sum of all samples, e.g.:

$$\triangle^{-1.2} x_k = 1x_k + 1.2x_{k-1} + 1.32x_{k-2} + 1.408x_{k-3} \ldots \qquad (42)$$

$$\triangle^{-0.8} x_k = 1x_k + 0.8x_{k-1} + 0.72x_{k-2} + 0.672x_{k-3} \ldots \qquad (43)$$

## 5. Diffusion approximation analysis of AQM performance

The presented mixed diffusion-simulation model uses the limited queue: G/G/1/N station to obtain the average queue length. Based on it, the simulation part of the model take a decision about dropping packets.

Calculations for both models (the mixed diffusion-simulation and simple simulation) were performed in Python and C. The simulations were done using the Simpy Python simulation packet. SimPy is released under the MIT License (free software license originating at the Massachusetts Institute of Technology). During the tests, we analyzed the following parameters of the transmission with AQM: the length of the queue, changes of the source intensity $\lambda$.

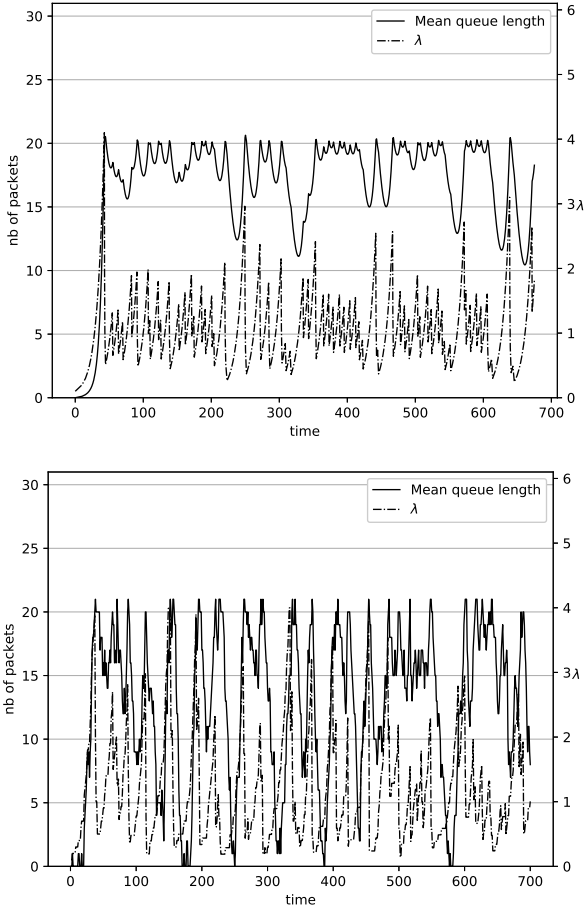

Figure 1: The router mean queue length and flow of the source, diffusion approximation (upper), simulation (bottom), RED algorithm.

The diffusion calculations were carried with $\Delta t = \frac{1}{\lambda}$ step. As a result, we obtain a queue distribution and then its mean size. On this basis of the simulation part of the model decides about the packet rejection. This decision influences the new value of $\lambda$. We consider delays $\Delta t$ related to get loss information by changes the TCP transmitter. Therefore the $\lambda$ reduction may occur not immediately but after a certain time $\Delta t$.

The information about the packet incoming is transferred to the AQM mechanism, which decides what to do with the given packet. As in the diffusion model, this decision affects the change in the intensity of the source. The results of the model were compared with pure simulation nodes.

The service time represented the time of a packet treatment and dispatching. Its distribution was exponential with constant parameter $\mu = 1$.

The task of the AQM controller is to make decisions about the incoming packet. AQM mechanism draws the value from range <0;1>. The drawn number is comparing with the value of the dropping packet function (this value depends on the queue length).

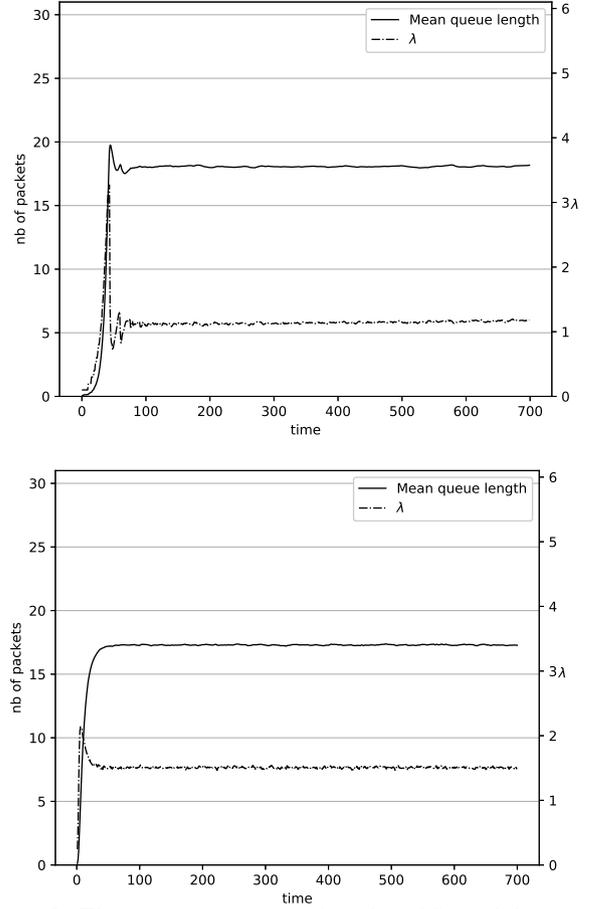

Figure 2: The router mean queue length and flow of the source, diffusion approximation (upper), simulation (bottom), RED algorithm.

We considered two types of AQM algorithms: standard RED and algorithm based on the answer of $PI^\alpha$ controller.

The queue parameters were as follows: maximal queue lengths = 30, RED parameters $Min_{th} = 10$, $Max_{th} = 20$, $PI^\alpha$ desire queue length = 10.

To obtain reliable transient state results, we repeated the experiments 10,000 times. An example of the results of a single experiment is shown in figure 1.

In the first stage of experiment, we considered the RED algorithm as AQM. Sample results of changes in the intensity of the queue's source and occupancy are presented in figure 2. Table 2 presents the detailed results for all performed experiments. It can be clearly seen that the results obtained by diffusion and simulations are similar.

In the experiments, we evaluate the AQM mechanism with packet dropping probability function based on $PI^\alpha$. We consider three different controllers. The table 1 presents the parameters of the used $PI^\alpha$ controllers.

Below we call a controller more powerful if there are more losses and the queue is shorter.

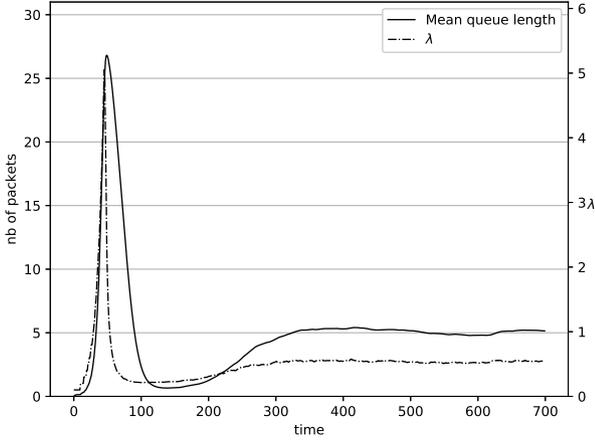

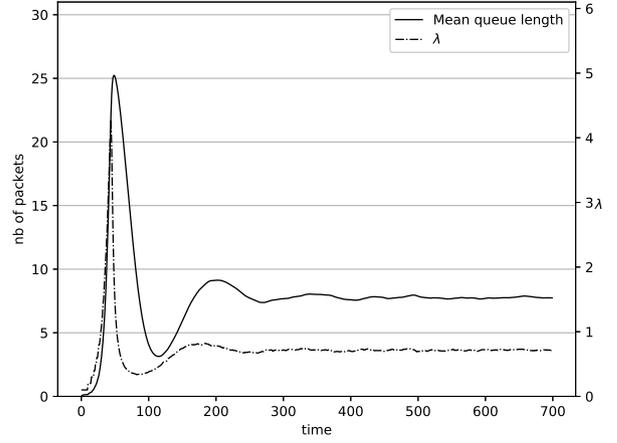

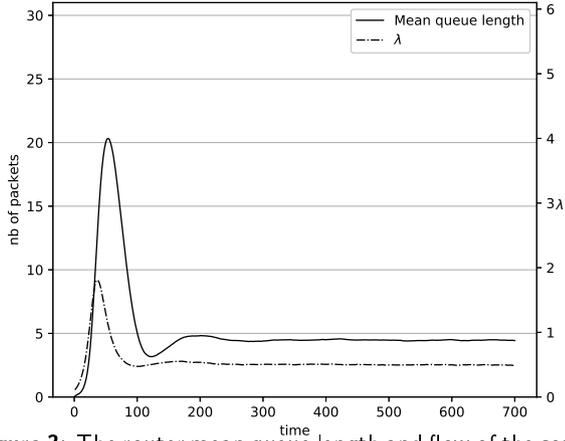

**Figure 3**: The router mean queue length and flow of the source, diffusion approximation (upper), simulation (bottom), 1st set of $PI^\alpha$ parameters.

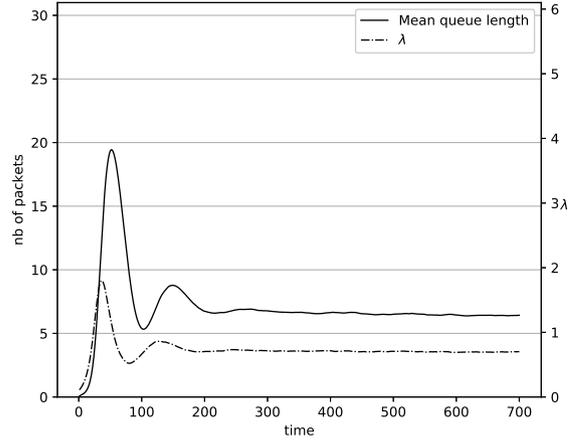

**Figure 4**: The router mean queue length and flow of the source, diffusion approximation (upper), simulation (bottom), 2nd set of $PI^\alpha$ parameters.

Figures 3, 4, 5 present how evaluates the intensity of the source over the time and how it affects the queue occupancy.

**Table 1**
$PI^\alpha$ controllers coefficients.

|   | $K_p$ | $K_i$ | $\alpha$ | setpoint |
|---|---|---|---|---|
| 1 | 0.0001 | 0.0004 | -1.2 | 10 |
| 2 | 0.0001 | 0.0014 | -0.8 | 10 |
| 3 | 0.0001 | 0.0040 | -0.4 | 10 |

**Table 2**
$PI^\alpha$ Obtained average queue lengths.

| AQM | Diffusion | Simulation |
|---|---|---|
| $RED$ | 18.067 | 17.106 |
| $PI^\alpha$ 1 | 5.100 | 5.251 |
| $PI^\alpha$ 2 | 7.780 | 7.123 |
| $PI^\alpha$ 3 | 10.431 | 10.075 |

In the next phase of the experiments, we evaluate the AQM mechanism with packet dropping probability function based on $PI^\alpha$. We consider three different controllers. The table 1 presents their parameters. The influence of the parameters on the efficiency of AQM queue was discussed in [12], [11], [13]. The proposed controllers differ in the effectiveness of maintaining the desired queue size.

The first controller is the most effective (we call it the most powerful) but it rejects a large number of packets. Inverse properties has the third controller. It rejects a small number of packages but allows significant deviations from the desired queue occupancy. The characteristics of the second controller are in between the properties of the first and the third controller.

Also as in the case of RED, differences in results obtained by diffusion approximation and simulations are negligible. However, the influence of controller parameters on the behavior of the AQM queue is clearly visible. The average queue length decreases with increasing controller power (table 2).

The advantage of diffusion approximation is the ability to consider two first moments of traffic (intensity and variation), instead of only one in case of more foreignful used fluid flow approximation. Figure 6 presents the influence of the traffic source variation on queue behavior. The differences become apparent only for great differences in variance.

We think that the reason for the small differences is the performance of TCP protocol and its mechanism of adapting the transmission speed to the network capabilities.

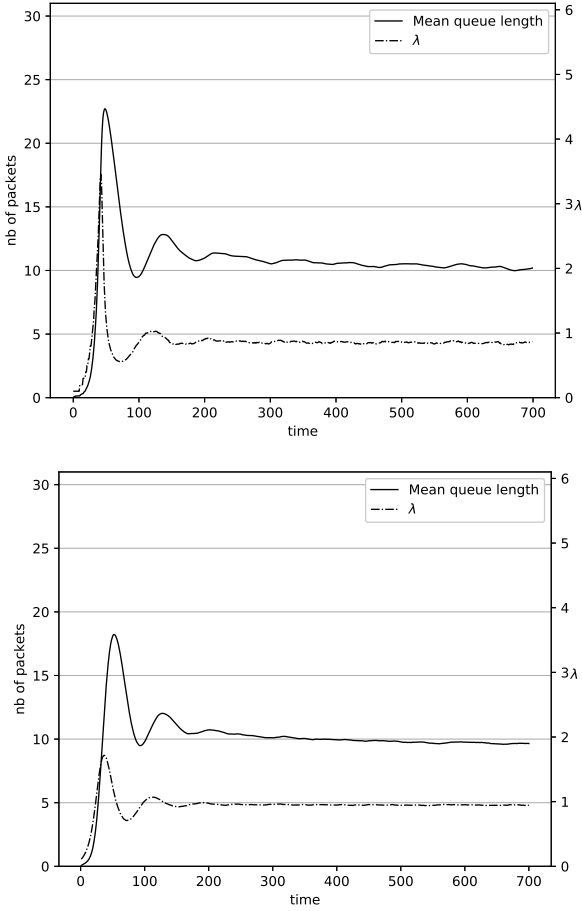

Figure 5: The router mean queue length and flow of the source, diffusion approximation (upper), simulation (bottom), 3rd set of $PI^\alpha$ parameters.

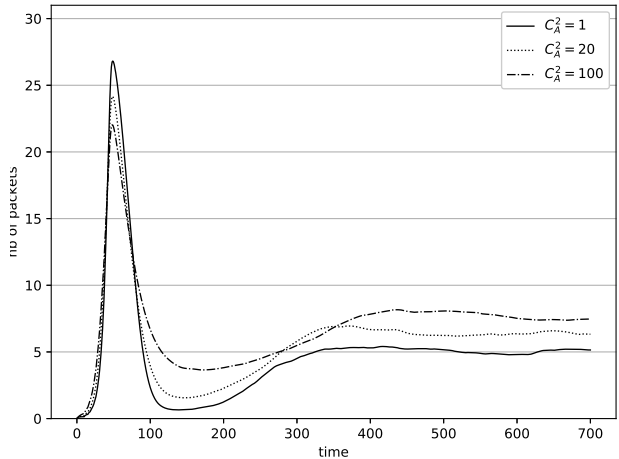

Figure 6: The router mean queue length, G/M/1/30 queue, impact of traffic variation, all sets of $PI^\alpha$ parameters.

## 6. Conclusions

This article describes the use of the diffusion approximation to estimate the behavior of AQM mechanisms and their influence on the evolution of the TCP congestion window. We proposed a simplified TCP NewReno transmitter model. According to it, the congestion window increases linearly by one packet of the successful transmission and decreases by half of its value in the case of a packet loss. In the case of diffusion approximation, the equivalent of the increase in CW is the increase in the intensity of the source $\lambda$ and an increase by one packet corresponds to an increase in lambda by an assumed value $\zeta$.

In our solution, we proposed combining the diffusion approximation with simulation. The approximation is used to calculate the size of the queue. The advantage of this approach is a natural description of transient states for any interarrival and service time distributions. The simulation part decides about a possible packet rejection (in accordance with the AQM rules). This decision affects the source intensity of $\lambda$. In our research, we considered the RED and $PI^\alpha$ algorithms.

We compare the results obtained from the proposed mixed diffusion-simulation method with the results obtained by simulation. In both cases, the results are consistent. In the case of the $PI^\alpha$ controller, the obtained average queue sizes are also correct and dependent on the controller's power. Increase in controller power reduces the average queue size. The results confirm the correctness of the proposed mixed model.

A certain disadvantage of the proposed model is the need to repeat the experiment numerous times in order to obtain reliable results. In the presented experiments, we have repeated the computations 10,000 times.

In our future work, we will focus on the diffusion models reflecting real Internet traffic (increased number of transmitters and the presence of both TCP and UDP streams).

## References

[1] Augustyn, D., Domański, A., Domańska, J.: A Choice of Optimal Packet Dropping Function for Active Queue Management. Communications in Computer and Information Science. Springer Berlin Heidelberg. vol. 79, pp. 199 – 206 (2010)

[2] Chang Feng, W., Kandlur, D., Saha, D.: Adaptive packet marking for maintaining end to end throughput in a differentiated service internet'. IEEE/ACM Transactions on Networking 7(5), pp. 685–697 (1999)

[3] Cox, R.P., Miller, H.D.: The Theory of Stochastic Processes. Chapman and Hall, London. (1965)

[4] Czachórski, T., Nycz, T., Pekergin, F.: Priority disciplines - A diffusion approach. 23rd International Symposium on Computer and Information Sciences, ISCIS. pp. 1 – 4 (2008)

[5] Czachórski, T., Nycz, T., Pekergin, F.: Transient States of Priority Queues - A Diffusion Approximation Study. Fifth Advanced International Conference on Telecommunications. pp. 44 – 51 (2009)

[6] Czachórski, T., Pekergin, F.: Diffusion Approximation as a Modelling Tool. Network Performance Engineering. A Handbook on Convergent Multi-Service Networks and Next Generation Internet. Springer Berlin Heidelberg. pp. 447 – 476 (2011)

[7] Domańska, J., Domański, A., Czachórski, T.: Fluid Flow Analysis of RED Algorithm with Modified Weighted Moving Average. Modern

Probabilistic Methods for Analysis of Telecommunication Networks. Springer Berlin Heidelberg. vol. 356, pp. 50 – 58 (2013)
[8] Domańska, J., Domański, A., Czachórski, T., Klamka, J.: The use of a non-integer order PI controller with an Active Queue Management Mechanism. International Journal of Applied Mathematics and Computer Science. vol. 26, pp. 777 – 789 (2016)
[9] Domańska, J., Domański, A., Czachórski, T., Klamka, J.: Self-similarity Trafic and AQM Mechanism Based on Non-integer Order $PI^\alpha D^\beta$ Controller. Communications in Computer and Information Science. Springer International Publishing. vol. 718, pp. 336 – 350 (2017)
[10] Domańska, J., Domański, A., Czachórski, T., Klamka, J., Marek, D., Szyguła, J.: GPU Accelerated Non-integer Order $PI^\alpha D^\beta$ Controller Used as AQM Mechanism. Communications in Computer and Information Science. Springer Verlag, Berlin. vol. 860, pp. 286 – 299 (2018)
[11] Domańska, J., Domański, A., Czachórski, T., Klamka, J., Marek, D., Szyguła, J.: The Influence of the Traffic Self-similarity on the Choice of the Non-integer Order $PI^\alpha$ Controller Parameters. Communications in Computer and Information Science. Springer International Publishing. vol. 935, pp. 76 – 83 (2018)
[12] Domańska, J., Domański, A., Czachórski, T., Klamka, J., Szyguła, J.: The AQM Dropping Packet Probability Function Based on Non-integer Order $PI^\alpha D^\beta$ Controller. Lecture Notes in Electrical Engineering. Springer International Publishing. vol. 496, pp. 36 – 48 (2019)
[13] Domańska, J., Domański, A., Czachórski, T., Klamka, J., Szyguła, J., Marek, D.: AQM mechanism with the dropping packet function based on the answer of several $PI^\alpha$ controllers. Communications in Computer and Information Science. Springer Verlag, Berlin. (2019)
[14] Domańska, J., Domański, A., Czachórski, T., Pagano, M.: The Fluid Flow Approximation of the TCP Vegas and Reno Congestion Control Mechanism. Computer and Information Sciences. Springer International Publishing. pp. 193 – 200 (2016)
[15] Floyd, S., Jacobson, V.: Random Early Detection gateways for congestion avoidance. IEEE/ACM Transactions on Networking. vol. 1(4), pp. 397 – 413 (1993)
[16] Gelenbe, E.: On Approximate Computer Systems Models. Journal of ACM. vol. 22(2), pp. 261 – 269 (1975)
[17] Halfin, S., Whitt, W.: Heavy-Traffic Limits for Queues with Many Exponential Servers. Operations Research. vol. 29, pp. 567 – 588 (1981)
[18] Hassan, M., Jain, R.: High Performance TCP/IP Networking - concepts, issues and solutions. Pearson Education Inc. (2004)
[19] Hollot, C., Misra, V., Towsley, D., Gong, W.: On designing improved controllers for AQM routers supporting TCP flows. Twentieth Annual Joint Conference of the IEEE Computer and Communications Society. Proceedings IEEE INFOCOM 2001. vol. 3, pp. 1726 – 1734 (2001)
[20] Iglehart, D.: Weak Convergence in Queueing Theory. Advances in Applied Probability. vol. 5(3), pp. 570 – 594 (1973)
[21] Iglehart, D., Whitt, W.: Multiple Channel Queues in Heavy Traffic - Part I. Advances in Applied Probability. vol. 2(1), pp. 150 – 177 (1970)
[22] Iglehart, D., Whitt, W.: Multiple Channel Queues in Heavy Trafic. Part II: Sequences, Networks, and Batches. Advances in Applied Probability. vol. 2(2), pp. 355 – 369 (1970)
[23] Kobayashi, H.: Application of the Diffusion Approximation to Queueing Networks I: Equilibrium Queue Distributions. Journal of the ACM (JACM). vol. 21(2), pp. 316 – 328 (1974)
[24] Kobayashi, H.: Modelling and Analysis: An Introduction to System Performance Evaluation Methodology. Addison-Wesley, Reading, Massachusetts. (1978)
[25] Krajewski, W., Viaro, U.: On robust fractional order PI controller for TCP packet flow. In: BOS Coference: Systems and Operational Research. Warsaw, Poland (September 2014)
[26] May, M., Bonald, T., Bolot, J.: Analytic evaluation of RED performance. In: Proceedings of the IEEE Infocom. Tel-Aviv, Izrael (2000)
[27] Melchor-Aquilar, D., Castillo-Tores, V.: Stability Analysis of Proportional-Integral AQM Controllers Supporting TCP Flows. Computacion y Sistemas. vol. 10, pp. 401 – 414 (2007)
[28] Melchor-Aquilar, D., Niculescu, S.: Computing non-fragile PI controllers for delay models of TCP/AQM networks. International Journal of Control. vol. 82(12), pp. 2249 – 2259 (2009)
[29] Michiels, W., Melchor-Aquilar, D., Niculescu, S.: Stability analysis of some classes of TCP/AQM networks. International Journal of Control, Manuscript. vol. 15, pp. 1 – 12 (2006)
[30] Newell, G.F.: Queues with time-dependent arrival rates. I — The transition through saturation. Journal of Applied Probability. vol. 2(2), pp. 436 – 451 (1968)
[31] Newell, G.F.: Queues with time-dependent arrival rates. II — The maximum queue and the return to equilibrium. Journal of Applied Probability. vol. 2(3), pp. 579 – 590 (1968)
[32] Newell, G.F.: Queues with time-dependent arrival rates. III — A mild rush hour. Journal of Applied Probability. vol. 2(3), pp. 591 – 606 (1968)
[33] Newell, G.F.: Applications of Queueing Theory. Monographs on Statistics and Applied Probability. Chapman and Hall, London (1971)
[34] Nonaka, Y., Nogami, S.: Evaluation of diffusion approximation for the G/G/1 queuing model. 8th Asia-Pacific Symposium on Information and Telecommunication Technologies. pp. 1 – 6 (2010)
[35] Podlubny, I.: Fractional order systems and $PI^\lambda d^\mu$ controllers. IEEE Transaction on Automatic Control. vol. 44(1), pp. 208 – 214 (1999)
[36] Quet, P., Ozbay, H.: On the design of AQM supporting TCP flows using robust control theory. IEEE Transactions on Automatic Control. vol. 49(6), pp. 1031 – 1036 (2004)
[37] Stehfest, H.: Algorithm 368: Numeric inversion of Laplace transform. Communications of the ACM, New York, USA. vol. 13(1), pp. 47 – 49 (1970)
[38] T.Czachórski: A method to solve Diffusion Equation with Instantaneous return Processes Acting as Boundary Conditions. Bulletin of Polish Academy of Sciences, Technical Sciences vol. 41(4), pp. 417 – 451 (1993)
[39] Ustebay, D., Ozbay, H.: Switching Resilient PI Controllers for Active Queue Management of TCP Flows. In: Proceedings of the 2007 IEEE International Conference on Networking, Sensing and Control. pp. 574 – 578 (2007)
[40] Whitt, W.: Multiple Channel Queues in Heavy Traffic - Part III: Random Server Selection. Advances in Applied Probability. vol. 2(2), pp. 370 – 375 (1970)
[41] Zheng, B., Atiquzzaman, M.: Dsred: A new queue management scheme for next generation networks. The 25th Annual IEEE Conference on Local Computer Networks pp. 242–251 (2000)
[42] Zheng, B., Atiquzzaman, M.: Improving Performance of Active Queue Management over Heterogeneous Networks. ICC 2001: International Conference on Communications. vol. 8, pp. 2375 – 2379 (2001)
[43] Zhou, K., Yeung, K., Li, V.: Nonlinear RED: A simple yet efficient Active Queue Management scheme. Computer Networks: The International Journal of Computer and Telecommunications Networking. vol. 50(18), pp. 3784 – 3794 (December 2006)

## A. My Appendix

## CRediT authorship contribution statement

**Dariusz Marek:** Conceptualization of this study, Methodology, Software, Writing - Original draft preparation. **Joanna Domańska:** Theory of Diffusion Approximation, Conceptualization of this study, Methodology, Software, Writing - Original draft preparation. **Tadeusz Czachórski:** Conceptualization of this study, Methodology, Software, Writing - Original draft preparation. **Jerzy Klamka:** Conceptualization of this study, Methodology, Software, Writing - Original draft preparation. **Jakub Szyguła:** Conceptualization of this study, Methodology, Software, Writing - Original draft preparation.

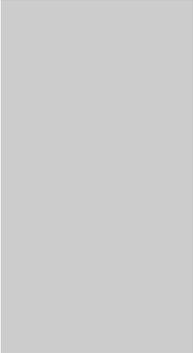

Dariusz Marek, Ph.D. Student in the Institute of Computer Science, Faculty of Automatic Control, Electronics and Computer Science, Silesian University of Technology.

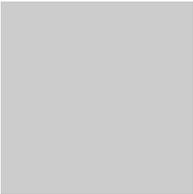

Adam Domański, Ph.D., Eng., works in the Computer Equipment Group of the Institute of Computer Science, Faculty of Automatic Control, Electronics and Computer Science, Silesian University of Technology. His main research interest in the computer networks domain is congestion control in packet networks.

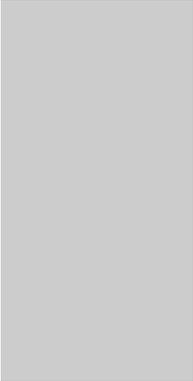

Joanna Domańska, Ph.D., Eng., works in the Computer Systems Modelling and Performance Evaluation Group of the Institute of Theoretical and Applied Informatics, Polish Academy of Sciences. Her main areas of research include performance modeling methods for computer networks.

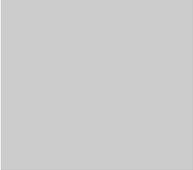

Tadeusz Czachórski, Ph.D., Prof., the head of the Institute of Theoretical and Applied Informatics of the Polish Academy of Sciences. His main areas of interest are mathematical and numerical methods for modeling and evaluation of computer networks.

Jerzy Klamka, Ph.D., Prof., a full member of the Polish Academy of Sciences, works in the Quantum Systems of Informatics Group of the Institute of Theoretical and Applied Informatics of the Polish Academy of Sciences. His main areas of research are controllability and observability of linear and nonlinear dynamical systems, and mathematical foundations of quantum computations. He is an author of monographs and numerous papers published in international journals.

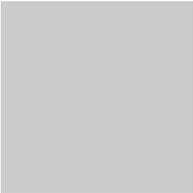

Jakub Szyguła, Ph.D. Student in the Institute of Computer Science, Faculty of Automatic Control, Electronics and Computer Science, Silesian University of Technology.